\documentclass{article}

\usepackage{arxiv}

\usepackage[utf8]{inputenc} % allow utf-8 input
\usepackage[T1]{fontenc}    % use 8-bit T1 fonts
\usepackage{hyperref}       % hyperlinks
\usepackage{url}            % simple URL typesetting
\usepackage{booktabs}       % professional-quality tables
\usepackage{amsmath}
\usepackage{amsfonts}       % blackboard math symbols
\usepackage{nicefrac}       % compact symbols for 1/2, etc.
\usepackage{microtype}      % microtypography
\usepackage{graphicx}
\usepackage{natbib}
\usepackage{doi}
\usepackage{tabularx}
\usepackage{xcolor}
\usepackage{colortbl}
\usepackage{multirow}
\usepackage{booktabs}
\usepackage{longtable}
\usepackage{float}
\usepackage{comment, soul}

\title{On Spatio-Temporal Stochastic Frontier Models}

\author{ \href{https://orcid.org/0000-0001-5976-1013}{\includegraphics[scale=0.06]{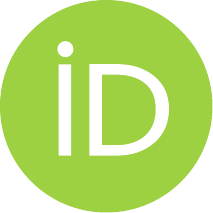}\hspace{1mm}Elisa Fusco\thanks{Corresponding author.}}\\
	Department of Statistics, Computer Science, Applications\\
	University of Florence\\
	Florence, Italy\\
	\texttt{elisa.fusco@unifi.it} \\
	\And
	\href{https://orcid.org/0000-0001-5173-3931}{\includegraphics[scale=0.06]{orcid.pdf}\hspace{1mm}Giuseppe Arbia} \\
	Department of Statistics\\
	Università Cattolica del Sacro Cuore\\
	Rome, Italy \\
	\texttt{giuseppe.arbia@unicatt.it} \\
	\And
	\href{https://orcid.org/0000-0003-4125-9337}{\includegraphics[scale=0.06]{orcid.pdf}\hspace{1mm}Francesco Vidoli} \\
	Department of Economics, Society and Politics \\
	University of Urbino\\
	Urbino, Italy \\
	\texttt{francesco.vidoli@uniurb.it} \\
 \And
	\href{https://orcid.org/0000-0002-7215-7934}{\includegraphics[scale=0.06]{orcid.pdf}\hspace{1mm}Vincenzo Nardelli} \\
	Università Cattolica del Sacro Cuore\\
	Rome, Italy \\
	\texttt{vincenzo.nardelli@unicatt.it} \\
}

\renewcommand{\shorttitle}{On Spatio-Temporal Stochastic Frontier Models}

\hypersetup{
pdftitle={On Spatio-Temporal Stochastic Frontier Models},
pdfsubject={stat.ME},
pdfauthor={Elisa~Fusco, Giuseppe~Arbia, Francesco~Vidoli, Vincenzo~Nardelli },
pdfkeywords={Stochastic frontier analysis, Spatio-temporal effects, Productive efficiency},
}

\begin{document}
\maketitle

\begin{abstract}
In the literature on stochastic frontier models until the early 2000s, the joint consideration of spatial and temporal dimensions was often inadequately addressed, if not completely neglected. However, from an evolutionary economics perspective, the production process of the decision-making units constantly changes over both dimensions: it is not stable over time due to managerial enhancements and/or internal or external shocks, and is influenced by the nearest territorial neighbours. This paper proposes an extension of the \cite{fusco2013spatial} SEM-like approach, which globally accounts for spatial and temporal effects in the term of inefficiency. In particular, coherently with the stochastic panel frontier literature, two different versions of the model are proposed: the time-invariant and the time-varying spatial stochastic frontier models. In order to evaluate the inferential properties of the proposed estimators, we first run Monte Carlo experiments and we then present the results of an application to a set of commonly referenced data, demonstrating robustness and stability of estimates across all scenarios.
\end{abstract}


\section{Introduction}
Understanding and analysing the efficiency of decision-making units has long been a cornerstone of different disciplines, including economics (\citealp{fare1994productivity}), agriculture (\citealp{kumbhakar1995efficiency}), and health studies (\citealp{hollingsworth2008measurement}). Traditionally, stochastic frontier analysis (SFA, \citealp{Aigner1977, battese1995model}) has served as a valuable tool for this purpose, providing information on both the level of production achieved and the potential for improvement. However, a crucial limitation of standard SFA lies in its neglect of the spatial context in which these units operate. Spatial dependence and spillover effects are increasingly recognised as significant factors that may influence production efficiency, which calls for analytical frameworks that incorporate these spatial dimensions.

This paper delves into the spatial stochastic frontier analysis (SSFA, \citealp{fusco2013spatial}) framework, an extension of SFA that explicitly accounts for spatial relationships and interactions. By acknowledging the spatial interconnections among decision-making units, SSFA provides a deeper and more detailed insight into efficiency differences. This enhanced framework allows us to disentangle the spatial aspects of inefficiency from pure unit-specific inefficiencies, revealing crucial insights into how location, proximity, and spatial interaction patterns influence production outcomes.\\
The rationale behind the use of SSFA is multifaceted. First, spatial dependence can arise from knowledge spillovers, shared infrastructure, or environmental features, leading to higher or lower efficiencies for neighbouring units depending on the nature of these interactions. Another point is that spatial autocorrelation in the error term could indicate the presence of unobserved spatial influences affecting production within the research region. Ignoring these factors might result in skewed assessments of technical efficiency and erroneous findings regarding the actual drivers of performance.
The SSFA method has been applied to estimate efficiency in various domains such as agriculture (\citealp{Vidolietal2016,Zulkarnain2020,Mittag2023}), local governments (\citealp{FuscoAllegrini20}), hospitals  (\citealp{Cavalieri2020}), food industry (\citealp{Cardamone2020}) or airports competition (\citealp{Bergantino2021}).

The extension of the SSFA to panel data is not simply an econometric and statistical exercise, but rather a response to a dynamic and evolutionary interpretation of economic firms, which operate in both time and space and undergo a continuous transition between equilibria. In fact, evolutionary approaches, which originated with \cite{darwin2016origin} and emphasised the mechanism of natural selection and the dynamic struggle for survival, found some early contributions to economics in \cite{Schumpeter1934} before being systematised by \cite{nelson1985evolutionary} and \cite{boschma1999evolutionary, Boschma2006}.\\
For evolutionists, three key processes take place in the marketplace: selection, variation, and reproduction, where essentially, in the struggle for survival, the species with the greatest ability to adapt to new conditions will survive with the greatest ability to adapt to the environment. Similarly, in the market, firms compete dynamically to attract consumers, and the market is therefore a selection mechanism for firms.
It shapes the opportunities and constraints on the growth, profitability, and likelihood of survival of firms, promotes "\emph{localized collective learning in a regional context}" and "\emph{the spatial formation of newly emerging industries as an evolutionary process}" (\citealp{boschma1999evolutionary}).\\
In this sense, dynamic efficiency (\emph{i.e.} the ability to innovate) is much more important than static (or allocative) efficiency, which makes it necessary to study economic processes over time. But the selection principle just seen does not imply that "selection" necessarily goes "from worst to best". The "best" and "worst" are concepts that depend on the specific selection mechanisms, their history, but also by the space in which they live, and the distribution of the characteristics actually present in a given ecology or market at a given time. And so it is also the ecological niche - read as a territorial or neighbour constraint - that conditions, in the long run, the mode of production, the efficiency, and hence the survival of firms. Therefore, it is because of these premises that the study of the economic efficiency of firms in both time and space is crucial, because the former shows the evolution of the system, while the latter ignores this dynamic, depending on whether or not they belong to an ecological niche\footnote{Clearly understood as neighbourhoods, both physically, but also as part of production chains, networks, districts.} in the market.

This paper aims to contribute to the growing body of knowledge on spatial stochastic frontier methods by presenting a new methodology for estimating production and cost frontiers for time-varying spatial data focussing on the composite error part. We achieve this by providing comprehensive evidence through simulations, analysis of real-world datasets, and the development of specific software code designed to implement our approach. Incorporating time-varying elements into spatial frontier models allows for a more nuanced understanding of how efficiency and productivity evolve in response to changing conditions and external factors.\\
We contribute to the existing literature in two distinct ways. First, we address a significant methodological gap by focussing on the most crucial aspect of the specification of stochastic frontiers: the composite error term, which encompasses both the random error and inefficiency components. By refining the modelling of the composite error, we improve the precision and interpretability of frontier estimates, enabling a more accurate assessment of inefficiency levels. Second, we extend this estimation framework to accommodate both time-varying and time-invariant cases, providing a comprehensive analysis that accounts for different types of temporal dynamics in spatial data and offering a versatile framework that can be adapted to a wide range of applications. 

The remainder of the paper is organised as follows. Section \ref{literature} introduces the spatial stochastic frontier models with a particular focus on dynamic models. Section \ref{method} introduces the proposed methodology for estimation in space and time, while Sections \ref{montecarlo} and \ref{riso} present, respectively, an application on simulated and real data. Section \ref{conclusion} discusses the main findings and concludes.

%%%%%%%%%%%%%%%%%%%
\section{A brief review on spatial stochastic frontier models for panel data}
\label{literature}

Research on frontier efficiency methods that simultaneously account for spatial and temporal effects with panel data is limited and started in the 2000s (for a detailed overview, see \citealp{Ayouba2023}).\\
The main problem, in fact, as mentioned by \cite{sfa_book2004} was that "\textit{in classical time invariant SFA the fixed effects (the $u_i$) are intended to capture the variation between producers in time-invariant technical efficiency. Unfortunately, they also capture the effects of all phenomena (such as the regulatory environment, for example) that vary across producers but that are time invariant for each producer}". 

The primary methods suggested in the literature over the past two decades to address this consideration can be categorized based on three issues: \emph{(i)} whether to consider efficiency as varying or constant over time; \emph{(ii)} the method for estimating efficiency, thus regression-based with a two-stage approach or maximum likelihood; \emph{(iii)} whether spatial dependence and spillover effects are only on the dependent variable, \emph{i.e.} the output (spatial lag specification - SAR), on the regressors (spatially lagged $X$ specification - SLX), on the error term (spatial Error model specification - SEM), on the first two (spatial Durbin specification - SDM), or on all terms (General nesting spatial specification - GNS).

%%%
A first stream of literature proposes the SAR frontier model assuming a time-variant, as in \cite{Glass2013, Glass2014} and \cite{Kutlu2019} or time-invariant, as in \cite{Han2016}, inefficiency term using a regression-based approach with a distribution-free approach to compute inefficiency in a second step.\\
Although distribution-free approaches have the advantage of not assuming a specific distribution for the inefficiency term, they are not robust to outliers. Therefore, \cite{Glass2016} proposed a SAR model for panel data integrating SAR and a half-normal stochastic frontier model proposed by \cite{Aigner1977} using a ML approach. Similarly, \cite{kutlu2020spatial} also employed the SAR approach in their spatial stochastic frontier analysis. \cite{Tsukamoto2019} extended the SAR stochastic frontier specification considering also the determinants of technical inefficiency, as proposed by \cite{Battese95} using a ML approach.

%%%%
A second group of scholars proposed a spatial Durbin specification as \cite{Adetutu15}, \cite{Glass2016}, \cite{Gude2018} and \cite{Ramajo2018} that integrates SLX and a half-normal stochastic frontier model; \cite{Galli2023} proposed a spatial Durbin stochastic frontier model for panel data introducing spillover effects in the determinants of technical efficiency (SDF-STE) using the ML approach.

%%%%
Finally, a third stream of panel data literature has been proposed including spatial dependence only in the error term or in the inefficiency term (see \emph{e.g.} \citealp{Druska04,Areal12,tsionas2016spatial}). In particular, \cite{Druska04} accounts for time-invariant fixed effects and uses a distribution-free approach, and \cite{Areal12}, \cite{tsionas2016spatial} uses a Bayesian approach. 

In our opinion, grasping spatial lag in this manner is essential, despite the fact that the term "spatial" is frequently employed in this specific literature without a clear definition: within an evolutionary reading of economic processes introduced earlier, in fact, it is necessary to separate the effects of the firm from those of the neighbourhood in terms of productive efficiency in order to take into account different evolutionary paths, neighbourhood, and/or spillover effects; in other words, in stochastic efficiency models, it is more important to develop spatial error models within composite error (error + inefficiency part) (thus generically borrowing spatial error (SEM) approaches) than to only try to describe more precisely the frontier (as the spatial lag models, SAR), assuming that it is homogeneous in space. In fact, there is no practical value in incorporating a generic $W$ on the frontier and then comparing all heterogeneous units with each other on a single frontier, but rather in showing how much space affects the lower or higher efficiency of each individual production unit.
Therefore, the composite error term lies at the heart of frontier models, and here the impact of the territory on the dynamic evolution of firms needs to be studied. 
%%%%%%%%%%%%%%%%%%%%%%%%%%%%%%%%%

\section{Space-time Stochastic Frontier Analysis (STSFA)}
\label{method}

As introduced previously, this paper proposes an extension of the \cite{fusco2013spatial, Fusco20SSFA} approach, which globally accounts for spatial and temporal effects in the term of inefficiency. In particular, coherently with the stochastic frontier literature (and specifically with \citealp{Battese1992}), this paper proposes two different versions of the model, namely: \emph{(i)} a time-invariant model and \emph{(ii)} a time-varying model. Notice that in both cases, the initial values for the maximum-likelihood estimation are derived from the algorithm. The two models will be discussed in detail in the next two sections.

%%%%%%%%%%
\subsection{The time-invariant STSFA}

In the proposed setting, we examine a production unit $i=1,...,N$ that, in a period $t=1,...,T$, uses $P$ inputs $\mathbf{x}_{it}=(x_{1it},...,x_{Pit})$, $\mathbf{x}_{it}\in \mathbb{R}_+^P$ to produce $Q$ outputs $\mathbf{y}_{it}=(y_{1it},...,y_{Qit})$, $\mathbf{y}_{it}\in \mathbb{R}_+^Q$. In this setting - and in the case of time-invariant - the generalisation of \cite{Fusco20SSFA} for panel data, \emph{i.e.} the \emph{Spatio-Temporal Stochastic Frontier Analysis} (STSFA), can be written as:\footnote{Note that as in \cite{fusco2013spatial, Fusco20SSFA} we consider the homoskedastic case and a Normal-Half-Normal distribution for the error term. Generalisations on these aspects can be developed in future research.}
\begin{equation}
\begin{split}
y_{it} = \; & \mathbf{x}_{it} \beta  + v_{it}  - s \cdot u_{i}  \\
 = \; & \mathbf{x}_{it}\beta  +v_{it} -  s \cdot  (\mathbf{I}-\rho \mathbf{W})^{-1}\widetilde{u}_{i}
\end{split}
\label{Model}
\end{equation}
where:
\begin{itemize}
\item $v_{it} \sim iid \ N(0,\sigma^{2}_{v}\mathbf{I})$;
\item $u_{i} \sim \ N^+(0,(\mathbf{I}-\rho \mathbf{W})^{-1}(\mathbf{I}-\rho  \mathbf{W}^T)^{-1}\sigma^{2}_{\widetilde{u}_{i}}\mathbf{I})$ is constant over the time;
\item  $u_i \ \text{and} \ v_{it}$ are independently distributed of each other, and of the regressors;
\item $\widetilde{u}_{i} \sim N(0,\sigma^{2}_{\widetilde{u}}\mathbf{I})$;
\item $s=1$ for production functions and $s=-1$ for cost functions.  
\end{itemize}

It is important to mention that in the proposed model, for the sake of simplicity, it has been assumed that the neighbourhood structure remains constant over time; in other terms, the spatial weight matrix $\mathbf{W}$ and the parameter $\rho$ have been set constant for all $t=1,...,T$.

The starting point for obtaining the general STSFA log-likelihood function, is to modify the density functions of $u$ and $v$ defined in \cite{Fusco20SSFA}\footnote{In order to simplify formulas $(1-\rho \sum_{i}w_{i.})$ is replaced by $\delta(\rho)$ as in \cite{Fusco20SSFA}.} accordingly with \cite{Battese1992}:
\begin{subequations}
\label{schema}
\begin{equation}
f([\delta(\rho)]^{-1}\widetilde{u}_{i})= \frac{2}{\sqrt{2 \pi} [\delta(\rho)]^{-1} \sigma_{\widetilde{u}}} \cdot exp\left \{ -\frac{[\delta(\rho)]^{-2}\widetilde{u}_{i}^2}{2[\delta(\rho)]^{-2}\sigma_{\widetilde{u}}^2} \right \} 
\label{densityU}
\end{equation}
\begin{equation}
f(\mathbf{v}_{i})= \frac{1}{(2\pi)^{T/2} \sigma_{v}^T} \cdot exp\left \{ -\frac{\mathbf{v}_{i}'\mathbf{v}_{i}}{2\sigma_{v}^2} \right \} 
\label{densityV}
\end{equation}
\end{subequations}
where $f([\delta(\rho)]^{-1}\widetilde{u}_{i})$ is the same of SSFA as $[\delta(\rho)]^{-1}\widetilde{u}_{i}$ is independent of time and $f(\mathbf{v}_{i})$ now is time dependent so $\mathbf{v}_{i}=(v_1,...,v_T)$ .

Given the independence assumption, the joint density function of $[\delta(\rho)]^{-1}\widetilde{u}_{i}$ and $\mathbf{v}_{i}$ can be written as:
\begin{equation}
f([\delta(\rho)]^{-1}\widetilde{u}_{i},\mathbf{v}_{i})= \frac{2}{(2\pi)^{(T+1)/2} [\delta(\rho)]^{-1} \sigma_{\widetilde{u}} \sigma_{v}^T} \cdot exp\left \{ -\frac{\mathbf{v}_{i}'\mathbf{v}_{i}}{2\sigma_{v}^2} -\frac{[\delta(\rho)]^{-2}\widetilde{u}_{i}^2}{2[\delta(\rho)]^{-2} \sigma_{\widetilde{u}}^2} \right \}
\label{JointDensityUV}
\end{equation}

Since in the general form $\epsilon_{it} = v_{it} - s \cdot [\delta(\rho)]^{-1} \widetilde{u}_{i}$; therefore the new joint density function for $[\delta(\rho)]^{-1}\widetilde{u}_{i}$ and $\epsilon_{i}=(v_1-s\cdot[\delta(\rho)]^{-1} \widetilde{u}_{i},...,v_T-s\cdot[\delta(\rho)]^{-1} \widetilde{u}_{i})$ becomes:
\begin{equation}
\begin{split}
f([\delta(\rho)]^{-1}\widetilde{u}_{i},\epsilon_{i})= & \frac{2}{(2\pi)^{(T+1)/2} [\delta(\rho)]^{-1} \sigma_{\widetilde{u}_i} \sigma_{v}^T} \cdot \\ & exp\left \{-\frac{ (s \cdot[\delta(\rho)]^{-1}\widetilde{u}_{i} - \mu_{*i})^2}{2\sigma^2_*} - \frac{\epsilon_{i}'\epsilon_{i}}{2\sigma^2_v} + \frac{\mu_{*i}^2}{2\sigma^2_*}\right \} \\
\ \\
\text{where}: \\
& \mu_{*i} = - \frac{T\overline{\epsilon}_{i} [\delta(\rho)]^{-2}\sigma_{\widetilde{u}}^2 }{\sigma_v^2 + T [\delta(\rho)]^{-2} \sigma_{\widetilde{u}}^2}, \quad \overline{\epsilon}_{i} = \frac{1}{T} \sum_{t=1}^{T} \epsilon_{it} \\
\ \\
& \sigma_{*}^2 =  \frac{[\delta(\rho)]^{-2} \sigma_{\widetilde{u}}^2 \sigma_v^2 }{\sigma_v^2 + T [\delta(\rho)]^{-2} \sigma_{\widetilde{u}}^2}
\end{split}
\label{JointDensityEV}
\end{equation}

and the marginal density function of $\epsilon_{i}$ is:\footnote{$\phi(\cdot)$ e $\Phi(\cdot)$ are the standard Normal density and distribution functions.}
\begin{equation}
\begin{split}
f(\epsilon_{i}) = & \int_{0}^{\infty } f([\delta(\rho)]^{-1}\widetilde{u}_{i},\epsilon_{i}) \ du \\
= & \; \frac{2[1-\Phi\left (-\mu_{*i}/\sigma_* \right )]}{(2\pi)^{T/2}\sigma_{v}^{T-1}(\sigma_v^2 + T [\delta(\rho)]^{-2} \sigma_{\widetilde{u}}^2)^{1/2} } \cdot exp\left \{ - \frac{\epsilon_{i}'\epsilon_{i}}{2\sigma^2_v} + \frac{\mu_{*i}^2}{2\sigma^2_*} \right\}\\
\end{split}
\label{DensityE}
\end{equation}
Whereupon, the sought general "time-invariant" log-likelihood function for the sample of $N$ producers and $T$ periods is given by:
\begin{equation}
\begin{split}
L = & costant - \frac{N(T-1)}{2} ln(\sigma_v^2)
- \frac{N}{2} ln(\sigma_v^2 + T [\delta(\rho)]^{-2} \sigma_{\widetilde{u}}^2) + \\
& + \sum_{i=1}^{N} ln \left [ 1-\Phi \left ( -\frac{\mu_{*i}}{\sigma*} \right ) \right ]  - \frac{1}{2}\sum_{i=1}^{N} \frac{\epsilon_{i}^2}{\sigma_v^2} + \frac{1}{2} \sum_{i=1}^{N} \left ( \frac{\mu_{*i}}{\sigma*}\right)^2
\end{split}
\label{Likelihood}
\end{equation}

Finally, the "time-invariant" technical efficiency of the firm $i$ at the time period $t$ is given by:
\begin{equation}
    TE_{it}= \left \{ \frac{1-\Phi\left [\sigma_* - (\mu_{*i}/\sigma_*) \right ]}{1-\Phi(-\mu_{*i}/\sigma_*)} \right \}\exp\left [-\mu_{*i}+\frac{1}{2} \sigma_*^{2}\right ]
\end{equation}
where $\mu_i^*$ and $\sigma^*$ are the same in Eq.(\ref{JointDensityEV}).

%%%%%%%%%%%%%%%%%%%%%%%%%%%%%%%%%
\subsection{The time varying STSFA}
In the case of time-varying assumption, the generalisation of \cite{Fusco20SSFA} for panel data, using the standard \cite{Battese1992} decay-model formulation, can be written as: 
\begin{equation}
\begin{split}
y_{it} = \; & \mathbf{x}_{it} \beta  + v_{it}  - s \cdot \eta_{it} \cdot u_{i}  \\
 = \; & \mathbf{x_{it}}\beta  +v_{it} -  s \cdot \exp \left\{-\eta(t-T) \right\} \cdot [(\mathbf{I}-\rho \mathbf{W})^{-1}\widetilde{u}_{i}]
\end{split}
\label{ModelTV}
\end{equation}

where $\eta_{it}=\exp \left\{-\eta(t-T) \right\}$ is a multiplicative factor, equal for all producers, that captures the dynamics of the degree of inefficiency. In particular, $\eta_{it}\geq 0$ decreases at an increasing rate if $\eta>0$, increases at an increasing rate if $\eta<0$, or remains constant if $\eta=0$.

Note that $u$ is once again independent of time, since the dynamics is given by $\eta_{it}$, so the density functions of $u$ and $v$ and the conjoint one are identical to Eqs. \eqref{densityU}, \eqref{densityV} and \eqref{JointDensityUV}. Instead, what is changing in time-varying formulation is $\epsilon_{it} = v_{it} - s \cdot \eta_{it} \cdot [\delta(\rho)]^{-1} \widetilde{u}_{i}$, therefore the new joint density function for $[\delta(\rho)]^{-1}\widetilde{u}_{i}$ and $\epsilon_{i}=(v_1-s\cdot\eta_{i1}\cdot[\delta(\rho)]^{-1} \widetilde{u}_{i},...,v_T-s\cdot\eta_{iT}\cdot[\delta(\rho)]^{-1} \widetilde{u}_{i})$ is:
\begin{equation}
\begin{split}
f([\delta(\rho)]^{-1}\widetilde{u}_{i},\epsilon_{i})= & \frac{2}{(2\pi)^{(T+1)/2} [\delta(\rho)]^{-1} \sigma_{\widetilde{u}_i} \sigma_{v}^T} \cdot \\ & exp\left \{-\frac{(s \cdot \eta_{i}[\delta(\rho)]^{-1}\widetilde{u}_{i} - \mu_{\eta i*})^2}{2\sigma_{\eta *}^2} - \frac{\epsilon_{i}'\epsilon_{i}}{2\sigma^2_v} + \frac{\mu_{\eta *i}^2}{2\sigma_{\eta *}^2}\right \} \\
\ \\
& \text{where}: \\
& \mu_{\eta i*} =  -\frac{\eta_i' \epsilon_{i}[\delta(\rho)]^{-2} \sigma_{\widetilde{u}}^2}{\sigma_v^2 + \eta_i'\eta_i[\delta(\rho)]^{-2} \sigma_{\widetilde{u}}^2 } \\
\ \\
& \sigma_{\eta*}^2 = \frac{[\delta(\rho)]^{-2} \sigma_{\widetilde{u}}^2 \sigma_v^2}{\sigma_v^2 + \eta'_i\eta_i[\delta(\rho)]^{-2} \sigma_{\widetilde{u}}^2 } \\
\end{split}
\label{JointDensityEV_TV}
\end{equation}
and $\eta_i$ represents the $T\times1$ vector of $\eta_{it}$'s associated with the time periods observed for the $i$-th firm.

The marginal density function of $\epsilon_{i}$ becomes:
\begin{equation}
\begin{split}
f(\epsilon_{i}) = & \int_{0}^{\infty } f([\delta(\rho)]^{-1}\widetilde{u}_{i},\epsilon_{i}) \ du \\
= & \; \frac{2[1-\Phi\left (-\mu_{\eta i*}/\sigma_{\eta*} \right )]}{(2\pi)^{T/2}\sigma_{v}^{T-1}(\sigma_v^2 + \eta'_i\eta_i [\delta(\rho)]^{-2} \sigma_{\widetilde{u}}^2) } \cdot exp\left \{ - \frac{\epsilon_{i}'\epsilon_{i}}{2\sigma^2_v} + \frac{\mu_{\eta i*}^2}{2\sigma_{\eta*}^2} \right\}\\
\end{split}
\label{DensityE_TV}
\end{equation}

The sought general "time-varying" log-likelihood function for the sample of $N$ producers and $T$ periods is:
\begin{equation}
\begin{split}
L = & costant - \frac{N(T-1)}{2} ln(\sigma_v^2) -  
 \frac{N}{2}\sum_{i=1}^{N} ln(\sigma_v^2 + \eta_i'\eta_i [\delta(\rho)]^{-2} \sigma_{\widetilde{u}}^2) \\
& + \sum_{i=1}^{N} ln \left [ 1-\Phi \left ( -\frac{\mu_{\eta i*}}{\sigma_{\eta*}} \right ) \right ]  - \frac{1}{2}\sum_{i=1}^{N} \frac{\epsilon_{i}^2}{\sigma_v^2} + \frac{1}{2} \sum_{i=1}^{N} \left ( \frac{\mu_{\eta i*}}{\sigma_{\eta*}}\right)^2
\ \\ 
\end{split}
\label{LikelihoodTV}
\end{equation} 

Finally, the "time-varying" technical efficiency of the firm $i$ at the time period $t$ can be derived from the formula:
\begin{equation}
    TE_{\eta it}= \left \{ \frac{1-\Phi\left [\sigma_{\eta*} - (\mu_{\eta i*}/\sigma_{\eta*}) \right ]}{1-\Phi(-\mu_{\eta i*}/\sigma_{\eta*})} \right \}\exp\left [-\mu_{\eta i*}+\frac{1}{2} \sigma_{\eta*}^2\right ]
\end{equation}
where $\mu_{\eta i*}$ and $\sigma_{\eta*}$ are the same in Eq.(\ref{JointDensityEV_TV}).

Note that if $\eta=0$, $\eta_i$ is equal to 1 and $\eta_i'\eta_i$ is equal to $T$, therefore, all equations collapse to their time-invariant versions.   

%%%%%%%%%%%%%%%%%%%%%%%%%%%%%%%%%
\section{A Monte Carlo evaluation of the STSFA model properties}
\label{montecarlo}

In order to evaluate the small sample inferential properties of the estimators proposed in the previous section, we will first conduct a thorough analysis by presenting the results of a series of Monte Carlo experiments. These experiments are designed to rigorously test the performance and robustness of the estimators under various conditions. 
The simulation process ensures that each generated sample reflects the specified characteristics and distributional assumptions, providing a robust framework to analyse the performance and estimation accuracy of various econometric models applied to stochastic frontier analysis. More specifically, the common DGP used to simulate 1,000 samples consisting of $100$, $200$, and $400$ observations for $T=5$ years is a single-input-single-output stochastic production frontier for panel data. This model is structured as follows:
\begin{equation}
y_{it} = 5 + 5\cdot x_i + v_i - u_{it}
\end{equation}
where:
\begin{itemize}
\item $x_i \sim Unif(0,1)$
\item $v_i \sim N(0.75\cdot(pi-2)/pi),1)$
\item $u_{it}= \exp \left\{-\eta(t-T) \right\} \cdot [(\mathbf{I}-\rho \mathbf{W})^{-1}\widetilde{u}_{i}]$ 
\item $\widetilde{u}_{i} \sim N^+(0,1)$.
\end{itemize}
It is important to highlight that for the purpose of emphasising the efficiency term in the simulation results, the regressor and the random term are kept constant throughout all 5 years. \\ Specifically, the frontier is estimated by the time-varying model for all combinations of $\rho=(0.05, 0.2, 0.4, 0.6, 0.8)$ and $\eta= (-0.10, -0.05, 0, 0.05, 0.10)$.\footnote{This sequence of values is chosen in order to increase or decrease the inefficiency by a maximum of 50\% and the value $0$ is included to test the time-invariant specific case.} \\ Finally, the spatial weight matrix is a sparse matrix, as suggested by \cite{Lesage09}, due to its low computational costs, where the average number of neighbours for each observation, identified by the nearest-neighbour method, is equal to $10\%$ of the number of observations $N$.

\begin{figure}[H]
\centering
\includegraphics[width=0.45\textwidth]{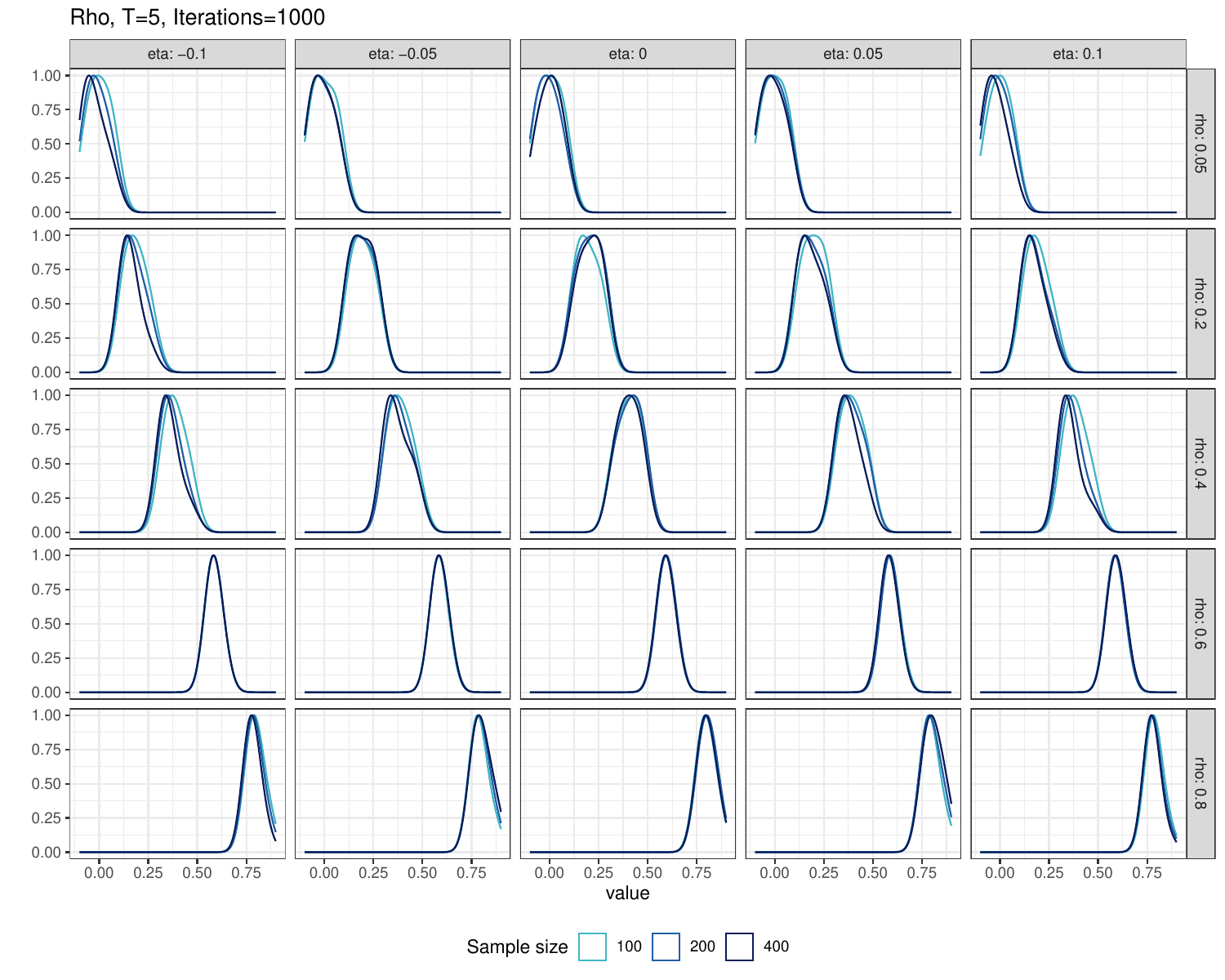}
\caption{Monte Carlo experiments results, $\rho$ parameter}
\label{fig:sim1}
\end{figure}

\begin{figure}[H]
\centering
\includegraphics[width=0.45\textwidth]{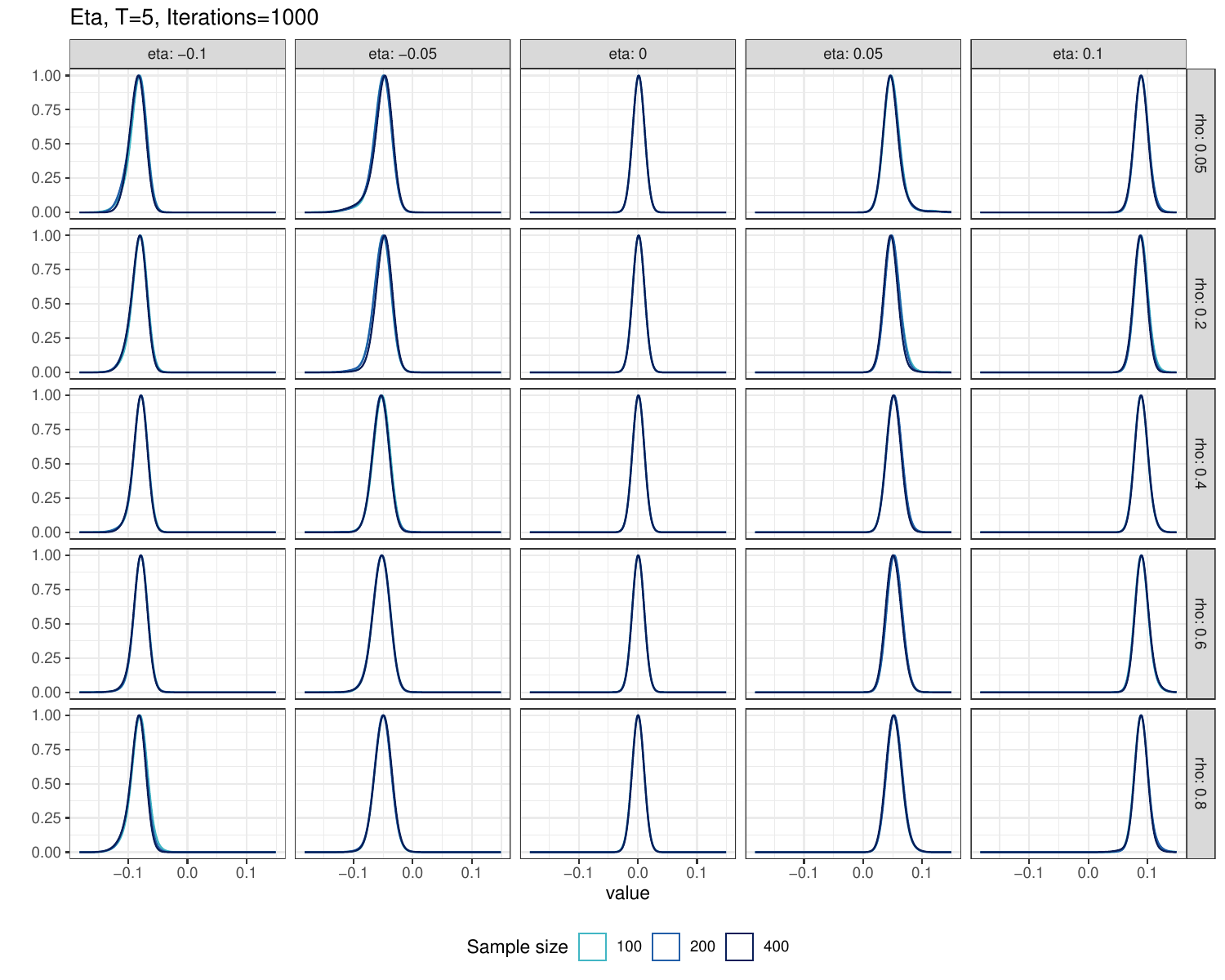}
\caption{Monte Carlo experiments results, $\eta$ parameter}
\label{fig:sim2}
\end{figure}

\begin{figure}[H]
\centering
\includegraphics[width=0.45\textwidth]{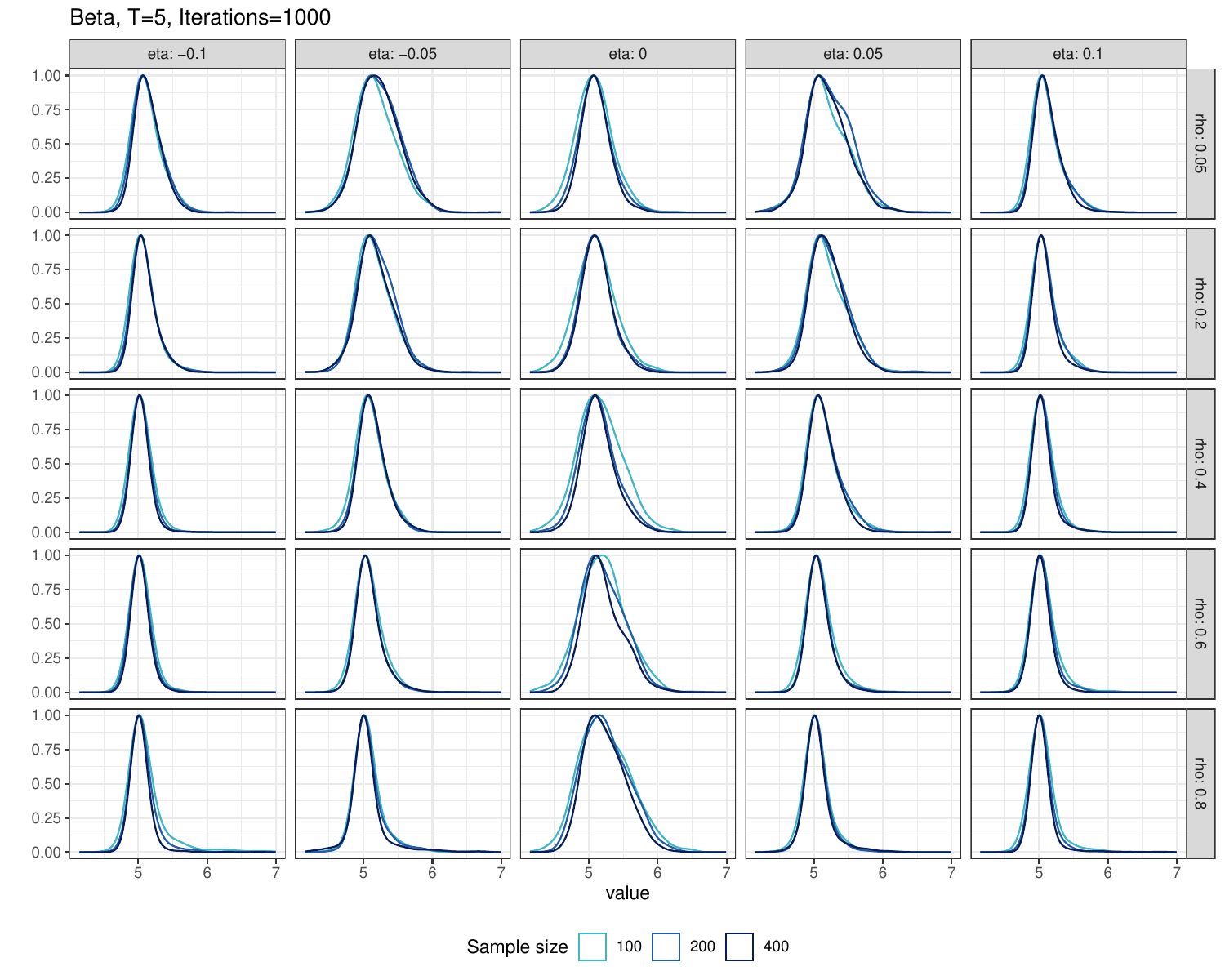}
\caption{Monte Carlo experiments results, $\beta$ parameter}
\label{fig:sim4}
\end{figure}

The parameters $\beta$, $\rho$ and $\eta$ are estimated according to the data generated by the rules and the selected DGP. The goodness of fit is assessed as usual by calculating the mean squared error (MSE) of the parameters and displaying the bias and standard deviation. The main findings are summarized as follows.\footnote{The elaborations have been carried out using the new functions of the \texttt{SSFA R} package (\citealp{ssfa_r}). The robustness check of the software is provided in \ref{pacchetto}.} The estimation of the parameters demonstrates very low bias, standard deviation and MSE, consistently across all parameters (Table \ref{tab:rmse400}). This precision improves as the sample size grows (see Tables \ref{tab:rmse100} and \ref{tab:rmse200} for 100 and 200 sample size results). Furthermore, the kernel densities in Figures \ref{fig:sim1} to \ref{fig:sim4}, illustrating that the distributions are closely aligned with the true values. 

\begin{table}[ht]
\centering
\caption{Monte Carlo experiments results (400 units)}
\resizebox{0.60\textwidth}{!}{
\begin{tabular}{crrrrrrr}
\\[-2ex]\hline 
\hline \\[-2ex]
\multirow{2}{*}{Parameter} & \multirow{2}{*}{$\eta$} & \multicolumn{6}{c}{$\rho$}\\[0.5ex] 
\cline{4-8} \\
[-2ex]
 & & & $0.05$ & $0.2$ & $0.4$ & $0.6$ & $0.8$ \\ 
\\[-2ex]\hline 
\hline \\[-2ex]
  \multirow{12}{*}{$\beta$} & \multirow{3}{*}{-0.1} 
& bias   & 0.0291 & 0.0218 & 0.0131 & 0.0098 & 0.0061 \\ 
&  & sd  & 0.0597 & 0.0578 & 0.0581 & 0.0578 & 0.0558 \\ 
&  & MSE & 0.0044 & 0.0038 & 0.0035 & 0.0034 & 0.0031 \\ 
\cline{4-8}
& \multirow{3}{*}{-0.05} 
& bias   & 0.0183 & 0.0302 & 0.0277 & 0.0132 & 0.0038 \\  
&  & sd  & 0.0719 & 0.0665 & 0.0652 & 0.0638 & 0.0595 \\  
&  & MSE & 0.0055 & 0.0053 & 0.0050 & 0.0042 & 0.0035 \\  
\cline{4-8}
& \multirow{3}{*}{0} 
& bias   & 0.0219 & 0.0199 & 0.0279 & 0.0295 & 0.0164 \\ 
&  & sd  & 0.0700 & 0.0719 & 0.0718 & 0.0718 & 0.0720 \\ 
&  & MSE & 0.0054 & 0.0056 & 0.0059 & 0.0060 & 0.0054 \\ 
\cline{4-8}
&  \multirow{3}{*}{0.05} 
& bias   & 0.0215 & 0.0249 & 0.0236 & 0.0165 & 0.0094 \\ 
&  & sd  & 0.0690 & 0.0684 & 0.0635 & 0.0619 & 0.0566 \\ 
&  & MSE & 0.0052 & 0.0053 & 0.0046 & 0.0041 & 0.0033 \\ 
\cline{4-8}
& \multirow{3}{*}{0.10} 
& bias   & 0.0260 & 0.0183 & 0.0131 & 0.0060 & 0.0028 \\  
&  & sd  & 0.0611 & 0.0561 & 0.0563 & 0.0593 & 0.0544 \\  
&  & MSE & 0.0044 & 0.0035 & 0.0033 & 0.0036 & 0.0030 \\  
   \cline{1-8}
   \multirow{12}{*}{$\rho$}  
&   \multirow{3}{*}{-0.10}
& bias   & -0.0385 & -0.0404 & -0.0393 & -0.0137 & -0.0131 \\ 
&  & sd  & 0.0426 & 0.0475 & 0.0513 & 0.0266 & 0.0370 \\ 
&  & MSE & 0.0033 & 0.0039 & 0.0042 & 0.0009 & 0.0015 \\ 
\cline{4-8}
& \multirow{3}{*}{-0.05} 
& bias   & -0.0301 & -0.0069 & -0.0257 & -0.0126 & 0.0226 \\ 
&  & sd  & 0.0441 & 0.0582 & 0.0597 & 0.0276 & 0.0594 \\ 
&  & MSE & 0.0028 & 0.0034 & 0.0042 & 0.0009 & 0.0040 \\ 
\cline{4-8}
& \multirow{3}{*}{0} 
& bias   & -0.0266 & 0.0071 & 0.0030 & -0.0066 & 0.0050 \\ 
&  & sd  & 0.0417 & 0.0576 & 0.0546 & 0.0284 & 0.0386 \\ 
&  & MSE & 0.0024 & 0.0033 & 0.0030 & 0.0008 & 0.0015 \\ 
\cline{4-8}
& \multirow{3}{*}{0.05} 
& bias   & -0.0296 & -0.0155 & -0.0241 & -0.0156 & 0.0249 \\ 
&  & sd  & 0.0431 & 0.0582 & 0.0534 & 0.0281 & 0.0561 \\ 
&  & MSE & 0.0027 & 0.0036 & 0.0034 & 0.0010 & 0.0038 \\ 
\cline{4-8}
& \multirow{3}{*}{0.10} 
& bias   & -0.0423 & -0.0274 & -0.0422 & -0.0071 & -0.0144 \\
&  & sd  & 0.0394 & 0.0521 & 0.0522 & 0.0287 & 0.0415 \\ 
&  & MSE & 0.0033 & 0.0035 & 0.0045 & 0.0009 & 0.0019 \\ 
   \cline{1-8}
  \multirow{12}{*}{$\eta$} 
   & \multirow{3}{*}{-0.1} 
   & bias & 0.0148 & 0.0165 & 0.0195 & 0.0196 & 0.0149 \\ 
&  & sd   & 0.0091 & 0.0095 & 0.0078 & 0.0078 & 0.0106 \\ 
&  & MSE  & 0.0003 & 0.0004 & 0.0004 & 0.0004 & 0.0003 \\ 
\cline{4-8}
& \multirow{3}{*}{-0.05} 
& bias   & -0.0019 & 0.0007 & -0.0040 & -0.0035 & -0.0001 \\
&  & sd  & 0.0154 & 0.0103 & 0.0094 & 0.0113 & 0.0106 \\ 
&  & MSE & 0.0002 & 0.0001 & 0.0001 & 0.0001 & 0.0001 \\ 
\cline{4-8}
& \multirow{3}{*}{0} 
& bias   & 0.0013 & 0.0012 & 0.0008 & 0.0005 & 0.0003 \\ 
&  & sd  & 0.0009 & 0.0009 & 0.0012 & 0.0014 & 0.0014 \\ 
&  & MSE & 0.0000 & 0.0000 & 0.0000 & 0.0000 & 0.0000 \\ 
\cline{4-8}
& \multirow{3}{*}{0.05} 
& bias   & -0.0013 & -0.0027 & 0.0019 & 0.0018 & 0.0026 \\ 
&  & sd  & 0.0122 & 0.0074 & 0.0072 & 0.0084 & 0.0093 \\ 
&  & MSE & 0.0002 & 0.0001 & 0.0001 & 0.0001 & 0.0001 \\ 
\cline{4-8}
& \multirow{3}{*}{0.10} 
& bias   & -0.0110 & -0.0124 & -0.0100 & -0.0086 & -0.0109 \\ 
&  & sd  & 0.0060 & 0.0050 & 0.0054 & 0.0068 & 0.0069 \\
&  & MSE & 0.0002 & 0.0002 & 0.0001 & 0.0001 & 0.0002 \\
\\[-1.8ex]\hline 
\hline \\[-1.8ex]
\end{tabular}}
\label{tab:rmse400}
\end{table}

%%%%%%%%%%%%%%%%%%%%%%%%%%%%%%%%%
\section{A re-examination of the popular Indonesian rice farm dataset (Feng and Horrace, 2012)}
\label{riso}

A well-known dataset used in the stochastic frontier literature to test our model is the Indonesian rice farm dataset published by \cite{Feng2012}. This dataset is particularly valuable due to its comprehensive nature, capturing a wide range of variables that influence rice production in Indonesia. It includes panel data with $N=171$ spatial observations across $T=6$ waves, allowing an in-depth analysis of temporal and spatial variations in farm efficiency. Detailed information on inputs such as labour, land, and capital, as well as outputs, provides a rich foundation for evaluating the performance and efficiency of rice farms, making it an excellent resource for modelling and testing theories in the context of agricultural productivity and efficiency. More specifically, the inputs to rice production included in the data set are seed (kg), urea (kg), trisodium phosphate (TSP) (kg), labour (hours of work) and land (hectares). The output is measured in kilogrammes of rice. The data also include dummy variables. DP equals 1 if pesticides were used and $0$ otherwise. DV1 equals 1 if high-yield rice varieties were planted and DV2 equals 1 if mixed varieties were planted. The DSS equals $1$ if it was a wet season, $0$ otherwise. The spatial weights matrix $W$ is constructed as in \cite{Druska04}: \emph{"Farms in the same village (out of six) are considered contiguous."}.

Results of STSFA time-invariant and time-varying models and a comparison with \cite{Druska04}\footnote{Results in "Table 3. Rice Regressions, Weighting Scheme M2 - Common Villages", on page 193 (\citealp{Druska04}).}
are reported in Table \ref{tab:ricefarm}. It is important to note that from a methodological point of view, there are some main differences between our method and that of \cite{Druska04}. In particular: \emph{(i)} in D\&H the model is time-invariant with fixed effects and the efficiency is estimated with a free-efficiency approach, instead in STSFA the model is time-invariant or time-varying and the efficiency is estimated with a ML approach; \emph{(ii)} in D\&H  all the error term is spatially lagged, instead in STSFA only the inefficiency term is spatially lagged; \emph{(iii)} in D\&H the $\rho$ parameter is estimated for each $t$ and then the average is kept instead in STSFA $\rho$ is global.

Main findings are that $\beta$ coefficients are close to \cite{Druska04}\footnote{Note that the robustness check of the time-invariant TSFA's intercept, $\sigma_u^2$ and $\sigma_v^2$ has been done by comparison with that of \cite{Horrace1996} (TSFA has been included in the Table \ref{tab:ricefarm} as baseline model). The only difference is the absence of some dummies.} but the use of fixed effects to account for inefficiency in the error term in the first step leads to some bias in H\&D's $\beta$ coefficients; our global $\rho$ is equal to 0.7103 (in a time-invariant model) and 0.7305 (in time-varying model) greater than D\&H due to the fact that all the error is lagged and perhaps the parameter can be lowered by the absence of spatial autocorrelation in the random term; very interesting the $\eta$ is positive in SFA and negative in time-varying STSFA. This might imply that disregarding the spatial dependence in the inefficiency term can lead to a biased $\eta$. Consequently, while the overall efficiency of farms increases over time as a result of positive externalities, individual farm efficiencies tend to decline over time.
% Table generated by Excel2LaTeX from sheet 'Foglio2'
\begin{table}[H]
  \centering
  \caption{Indonesian rice farm estimation results by method}
  \resizebox{0.6\textwidth}{!}{
    \begin{tabular}{llll|ll}
    \\[-2ex]\hline 
\hline \\[-2.5ex]
          & \multicolumn{3}{c|}{Time-invariant} & \multicolumn{2}{c}{Time-varying} \\
          \cline{2-6}\\[-2.5ex]
          & \multicolumn{1}{c}{TSFA} & \multicolumn{1}{c}{STSFA} & \multicolumn{1}{c|}{D\&H} & \multicolumn{1}{c}{TSFA} & \multicolumn{1}{c}{STSFA} \\
\\[-2.5ex]\hline 
\hline \\[-2.5ex]
    Intercept  & \multicolumn{1}{l}{$5.1087^{ *** }$} & \multicolumn{1}{l}{$5.5680^{ *** }$} & -     & \multicolumn{1}{l}{$5.3182^{ *** }$} & \multicolumn{1}{l}{$5.5869^{ ***}$} \\
      Seed  & \multicolumn{1}{l}{$0.1430^{ *** }$} & \multicolumn{1}{l}{$0.1469^{ *** }$}  & $0.1035^{*}$ & \multicolumn{1}{l}{$0.1368^{ *** }$} & \multicolumn{1}{l}{$0.1421^{ ***}$} \\
      Urea  & \multicolumn{1}{l}{$0.1112^{ *** }$} & \multicolumn{1}{l}{$0.0852^{ *** }$}  & $0.0909^{*}$ & \multicolumn{1}{l}{$0.1053^{ *** }$} & \multicolumn{1}{l}{$0.0896^{ ***}$} \\
      TSP  & \multicolumn{1}{l}{$0.0781^{ *** }$} & \multicolumn{1}{l}{$0.0813^{ *** }$} & $0.0356^{*}$ & \multicolumn{1}{l}{$0.0782^{ *** }$} & \multicolumn{1}{l}{$0.0797^{ ***}$} \\
      Labor  & \multicolumn{1}{l}{$0.2286^{ *** }$} & \multicolumn{1}{l}{$0.2110^{ *** }$} & $0.2385^{*}$ & \multicolumn{1}{l}{$0.2054^{ *** }$} & \multicolumn{1}{l}{$0.2117^{ ***}$} \\
      Land  & \multicolumn{1}{l}{$0.4692^{ *** }$} & \multicolumn{1}{l}{$0.4947^{ *** }$} & $0.4855^{*}$ & \multicolumn{1}{l}{$0.4995^{ *** }$} & \multicolumn{1}{l}{$0.4948^{ ***}$} \\
      DP  & \multicolumn{1}{l}{$0.0156^{   }$} & \multicolumn{1}{l}{$0.0098^{ *** }$} & $-0.0189$ & \multicolumn{1}{l}{$0.0157^{   }$} & \multicolumn{1}{l}{$0.0124^{  }$} \\
      DV1  & \multicolumn{1}{l}{$0.1615^{ *** }$} & \multicolumn{1}{l}{$0.1796^{   }$} & $0.1116^{*}$ & \multicolumn{1}{l}{$0.1578^{ *** }$} & \multicolumn{1}{l}{$0.1822^{ ***}$} \\
      DV2  & \multicolumn{1}{l}{$0.1349^{ ** }$} & \multicolumn{1}{l}{$0.1304^{ *** }$} & $0.1080^{*}$ & \multicolumn{1}{l}{$0.1289^{ * }$} & \multicolumn{1}{l}{$0.1359^{ **}$} \\
      DSS  & \multicolumn{1}{l}{$0.0473^{ * }$} & \multicolumn{1}{l}{$0.0507^{ * }$} & $0.0789$ & \multicolumn{1}{l}{$0.0509^{ * }$} & \multicolumn{1}{l}{$0.0505^{ *}$} \\
      $\sigma^2_{u_{dmu}}$  & \multicolumn{1}{l}{-} & 0.0001 & -     & \multicolumn{1}{l}{-} & 0.001 \\
      $\sigma^2_{u_{sar}}$  & \multicolumn{1}{l}{-} & 0.0186  & -     & \multicolumn{1}{l}{-} & 0.0184 \\
      $\sigma^2_{u}$  & \multicolumn{1}{l}{$0.0209^{ * }$} & \multicolumn{1}{l}{-}  & -     & \multicolumn{1}{l}{$0.0183^{ * }$} & \multicolumn{1}{l}{-} \\
      $\sigma^2_{v}$  & \multicolumn{1}{l}{$0.1099^{ *** }$} & \multicolumn{1}{l}{$0.1081^{ *** }$}  & -     & \multicolumn{1}{l}{$0.1095^{ *** }$} & \multicolumn{1}{l}{$0.1086^{ ***}$} \\
      $\rho$  & \multicolumn{1}{l}{-} & 0.7103 &  \multicolumn{1}{r|}{0.6604} & \multicolumn{1}{l}{-} & 0.7305 \\
      $\eta$  & \multicolumn{1}{l}{-} & \multicolumn{1}{l}{-} &    -   & 0.0376 & -0.0110 \\
      AIC  & -921.6 & -948.48 & -     & -919.33 & -945.50 \\
      \\[-2ex]\hline 
\hline \\[-2ex]
    \end{tabular}%
    }
  \label{tab:ricefarm}%
\end{table}%

%%%%%%%%%%%%%%%%%%%%%%%%%%%%%%%%%
\section{Final remarks}
\label{conclusion}

Economic processes are increasingly viewed as dynamic and constantly evolving and adapting, often influenced by spatial interactions and externalities, such as technology spillovers, local innovation, and resource sharing among neighbouring regions. \\
The introduction of the STSFA family of estimators marks a significant advancement in evolutionary economics by incorporating both spatial and temporal dependencies into the stochastic frontier model and focussing on the composite error term. Traditional models, in fact, often fail to account for these factors, leading to biased efficiency estimates. In particular, neglecting temporal dynamics can cause inefficiencies to be underestimated or overestimated. By integrating spatial and temporal dynamics, the STSFA model offers a more nuanced and accurate analysis of efficiency, which is crucial to understanding complex production processes. The STSFA model complements other methods proposed in the literature, but differs from them in that it focusses on modelling the error and inefficiency part, which we believe is precisely the core of frontier models; according to our perspective, in fact, it is more beneficial to understand the reasons and locations of systematic deviations from efficient behaviour by firms rather than focussing on the frontier itself.\\
This deeper insight is especially useful for policymakers who aim to create strategies that boost productivity. The STSFA model provides a sophisticated framework for identifying regions where policy interventions can have the greatest impact. By capturing the interaction between spatial and temporal factors, the model enables the identification of areas that would benefit the most from targeted policy measures. This information is vital for designing policies that not only improve efficiency, but also promote sustainable development.\\
To support the estimation of the STSFA model, the SSFA \texttt{R} package has been updated, offering a robust tool for researchers and facilitating a deeper exploration of efficiency in different sectors and geographic areas. \\
Finally, further applications of the model across various sectors and geographical areas would be valuable in generalising the findings, also exploring non-linear spatial dependencies and other forms of interactions between units.

%%%%%%%%%%%%%%%%%%%%%%%%%%%%%%%%%

\bibliographystyle{unsrtnat}
\bibliography{main}

%%%%%%%%%%%%%%%%%%%%%%%%%%%%%%%%%

\newpage
\appendix
%\setcounter{figure}{0}
%\setcounter{table}{0}

%%%%%%%%%%%
\section{SSFA R package - robustness check}
\label{pacchetto}
The proposed STSFA time-invariant and time-varying models have been implemented in two new functions of the \texttt{SSFA R} package by \cite{ssfa_r}. \texttt{SSFA} package allows to estimate six different model specifications: SFA and SSFA cross-section, TSFA and STSFA time-invariant and TSFA and STSFA time-varying.

With the aim of testing the robustness of the new functions, a comparison of the time-invariant and time-varying TSFA models (subsequently modified to account for spatial autocorrelation) with the Fortran source code of Frontier 4.1 (\texttt{frontier R} package by \citealp{frontier_r}) has been carried out, obtaining almost identical results as shown in Table \ref{tab:soft_check}.\footnote{Coelli, T. (1996) A Guide to FRONTIER Version 4.1: A Computer Program for Stochastic Frontier Production and Cost Function Estimation, CEPA Working Paper 96/08, \url{http://www.uq.edu.au/economics/cepa/frontier.php}, University of New England.}
% latex table generated in R 4.3.2 by xtable 1.8-4 package
% Thu Jun 20 21:02:09 2024
\begin{table}[ht]
\centering
\caption{SSFA package robustness check on classical SFA for panel data}
\resizebox{0.45\textwidth}{!}{
\begin{tabular}{lrrrr}
\\[-1.8ex]\hline 
\hline \\[-1.8ex]
& \multicolumn{2}{c}{Time-invariant} & \multicolumn{2}{c}{Time-varying} \\
 & \texttt{SSFA}  & \texttt{frontier} & \texttt{SSFA} & \texttt{frontier} \\ 
\\[-1.8ex]\hline 
\hline \\[-1.8ex]
Intercept & 5.1087 & 5.1027 & 5.3182 & 5.1193 \\ 
  Seed & 0.1430 & 0.1425 & 0.1368 & 0.1428 \\ 
  Urea & 0.1112 & 0.1114 & 0.1053 & 0.1106 \\ 
  TSP & 0.0781 & 0.0778 & 0.0782 & 0.0762 \\ 
  Labor & 0.2286 & 0.2297 & 0.2054 & 0.2285 \\ 
  Land & 0.4692 & 0.4687 & 0.4995 & 0.4716 \\ 
  DP & 0.0156 & 0.0157 & 0.0157 & 0.0177 \\ 
  DV1 & 0.1615 & 0.1617 & 0.1578 & 0.1584 \\ 
  DV2 & 0.1349 & 0.1327 & 0.1289 & 0.1294 \\ 
  DSS & 0.0473 & 0.0467 & 0.0509 & 0.0511 \\ 
  \midrule
  $\sigma^2$ & 0.1308 & 0.1307 & 0.1278 & 0.1273 \\ 
  $\eta$ & - & - & 0.0376 & 0.0367 \\ 
\\[-1.8ex]\hline 
\hline \\[-1.8ex]
\end{tabular}}
\label{tab:soft_check}
\end{table}

%%%%%%%%%%%
\newpage
\section{RMSE results by different simulation setting}
\label{RMSE_simula}
\begin{table}[H]
\centering
\caption{Monte Carlo experiments results (100 units)}
\resizebox{0.65\textwidth}{!}{
\begin{tabular}{crrrrrrr}
\\[-2ex]\hline 
\hline \\[-2ex]
\multirow{2}{*}{Parameter} & \multirow{2}{*}{$\eta$} & \multicolumn{6}{c}{$\rho$}\\[0.5ex] 
\cline{4-8} \\
[-2ex]
 & & & $0.05$ & $0.2$ & $0.4$ & $0.6$ & $0.8$ \\ 
\\[-2ex]\hline 
\hline \\[-2ex]
  \multirow{12}{*}{$\beta$} & \multirow{3}{*}{-0.1} 
& bias   & 0.0238 & 0.0164 & 0.0110 & 0.0092 & 0.0115 \\  
&  & sd  & 0.0680 & 0.0692 & 0.0674 & 0.0683 & 0.0691 \\  
&  & MSE & 0.0052 & 0.0051 & 0.0046 & 0.0047 & 0.0049 \\  
\cline{4-8}
& \multirow{3}{*}{-0.05} 
& bias   & 0.0145 & 0.0137 & 0.0195 & 0.0143 & 0.0095 \\  
&  & sd  & 0.0755 & 0.0723 & 0.0708 & 0.0680 & 0.0657 \\  
&  & MSE & 0.0059 & 0.0054 & 0.0054 & 0.0048 & 0.0044 \\  
\cline{4-8}
& \multirow{3}{*}{0} 
& bias   & 0.0109 & 0.0225 & 0.0135 & 0.0196 & 0.0190 \\ 
&  & sd  & 0.0751 & 0.0748 & 0.0723 & 0.0734 & 0.0760 \\ 
&  & MSE & 0.0058 & 0.0061 & 0.0054 & 0.0058 & 0.0061 \\ 
\cline{4-8}
&  \multirow{3}{*}{0.05} 
& bias   & 0.0211 & 0.0254 & 0.0197 & 0.0178 & 0.0091 \\ 
&  & sd  & 0.0700 & 0.0711 & 0.0701 & 0.0690 & 0.0669 \\ 
&  & MSE & 0.0053 & 0.0057 & 0.0053 & 0.0051 & 0.0045 \\ 
\cline{4-8}
& \multirow{3}{*}{0.10} 
& bias   & 0.0154 & 0.0161 & 0.0136 & 0.0116 & 0.0103 \\ 
&  & sd  & 0.0681 & 0.0686 & 0.0676 & 0.0678 & 0.0665 \\ 
&  & MSE & 0.0049 & 0.0050 & 0.0048 & 0.0047 & 0.0045 \\ 
   \cline{1-8}
   \multirow{12}{*}{$\rho$}  
&   \multirow{3}{*}{-0.10}
& bias   & -0.0287 & -0.0134 & -0.0111 & -0.0133 & 0.0039 \\
&  & sd  & 0.0446 & 0.0542 & 0.0527 & 0.0258 & 0.0425 \\ 
&  & MSE & 0.0028 & 0.0031 & 0.0029 & 0.0008 & 0.0018 \\ 
\cline{4-8}
& \multirow{3}{*}{-0.05} 
& bias   & -0.0255 & -0.0055 & -0.0119 & -0.0141 & 0.0027 \\
&  & sd  & 0.0461 & 0.0551 & 0.0574 & 0.0264 & 0.0499 \\ 
&  & MSE & 0.0028 & 0.0031 & 0.0034 & 0.0009 & 0.0025 \\ 
\cline{4-8}
& \multirow{3}{*}{0} 
& bias   & -0.0228 & -0.0061 & 0.0051 & -0.0061 & 0.0030 \\ 
&  & sd  & 0.0460 & 0.0567 & 0.0571 & 0.0283 & 0.0384 \\ 
&  & MSE & 0.0026 & 0.0032 & 0.0033 & 0.0008 & 0.0015 \\ 
\cline{4-8}
& \multirow{3}{*}{0.05} 
& bias   & -0.0250 & -0.0010 & -0.0044 & -0.0103 & 0.0016 \\ 
&  & sd  & 0.0437 & 0.0578 & 0.0575 & 0.0289 & 0.0435 \\ 
&  & MSE & 0.0025 & 0.0033 & 0.0033 & 0.0009 & 0.0019 \\ 
\cline{4-8}
& \multirow{3}{*}{0.10} 
& bias   & -0.0296 & -0.0102 & -0.0108 & -0.0102 & -0.0058 \\ 
&  & sd  & 0.0412 & 0.0546 & 0.0542 & 0.0266 & 0.0397 \\ 
&  & MSE & 0.0026 & 0.0031 & 0.0031 & 0.0008 & 0.0016 \\ 
   \cline{1-8}
  \multirow{12}{*}{$\eta$} 
   & \multirow{3}{*}{-0.1} 
   & bias & 0.0157 & 0.0180 & 0.0199 & 0.0203 & 0.0181 \\ 
&  & sd   & 0.0116 & 0.0101 & 0.0093 & 0.0084 & 0.0117 \\ 
&  & MSE  & 0.0004 & 0.0004 & 0.0005 & 0.0005 & 0.0005 \\ 
\cline{4-8}
& \multirow{3}{*}{-0.05} 
& bias   & -0.0028 & -0.0021 & -0.0030 & -0.0033 & -0.0005 \\ 
&  & sd  & 0.0146 & 0.0116 & 0.0104 & 0.0112 & 0.0104 \\ 
&  & MSE & 0.0002 & 0.0001 & 0.0001 & 0.0001 & 0.0001 \\ 
\cline{4-8}
& \multirow{3}{*}{0} 
& bias   & 0.0013 & 0.0011 & 0.0008 & 0.0004 & 0.0002 \\ 
&  & sd  & 0.0011 & 0.0015 & 0.0019 & 0.0022 & 0.0028 \\ 
&  & MSE & 0.0000 & 0.0000 & 0.0000 & 0.0000 & 0.0000 \\ 
\cline{4-8}
& \multirow{3}{*}{0.05} 
& bias   & -0.0003 & -0.0000 & 0.0026 & 0.0031 & 0.0028 \\ 
&  & sd  & 0.0122 & 0.0096 & 0.0082 & 0.0080 & 0.0088 \\ 
&  & MSE & 0.0001 & 0.0001 & 0.0001 & 0.0001 & 0.0001 \\ 
\cline{4-8}
& \multirow{3}{*}{0.10} 
& bias   & -0.0097 & -0.0102 & -0.0105 & -0.0097 & -0.0101 \\ 
&  & sd  & 0.0066 & 0.0064 & 0.0057 & 0.0068 & 0.0073 \\ 
&  & MSE & 0.0001 & 0.0001 & 0.0001 & 0.0001 & 0.0002 \\ 
\\[-1.8ex]\hline 
\hline \\[-1.8ex]
\end{tabular}}
\label{tab:rmse100}
\end{table}
\newpage
\begin{table}[H]
\centering
\caption{Monte Carlo experiments results (200 units)}
\resizebox{0.65\textwidth}{!}{
\begin{tabular}{crrrrrrr}
\\[-2ex]\hline 
\hline \\[-2ex]
\multirow{2}{*}{Parameter} & \multirow{2}{*}{$\eta$} & \multicolumn{6}{c}{$\rho$}\\[0.5ex] 
\cline{4-8} \\
[-2ex]
 & & & $0.05$ & $0.2$ & $0.4$ & $0.6$ & $0.8$ \\ 
\\[-2ex]\hline 
\hline \\[-2ex]
  \multirow{12}{*}{$\beta$} & \multirow{3}{*}{-0.1} 
& bias   & 0.0247 & 0.0196 & 0.0147 & 0.0093 & 0.0075 \\ 
&  & sd  & 0.0653 & 0.0627 & 0.0626 & 0.0643 & 0.0596 \\ 
&  & MSE & 0.0049 & 0.0043 & 0.0041 & 0.0042 & 0.0036 \\ 
\cline{4-8}
& \multirow{3}{*}{-0.05} 
& bias   & 0.0237 & 0.0253 & 0.0312 & 0.0156 & 0.0084 \\  
&  & sd  & 0.0709 & 0.0708 & 0.0661 & 0.0671 & 0.0620 \\  
&  & MSE & 0.0056 & 0.0056 & 0.0053 & 0.0047 & 0.0039 \\  
\cline{4-8}
& \multirow{3}{*}{0} 
& bias   & 0.0229 & 0.0175 & 0.0193 & 0.0159 & 0.0186 \\  
&  & sd  & 0.0732 & 0.0705 & 0.0726 & 0.0720 & 0.0746 \\  
&  & MSE & 0.0059 & 0.0053 & 0.0056 & 0.0054 & 0.0059 \\  
\cline{4-8}
&  \multirow{3}{*}{0.05} 
& bias   & 0.0185 & 0.0257 & 0.0255 & 0.0204 & 0.0116 \\ 
&  & sd  & 0.0712 & 0.0705 & 0.0664 & 0.0645 & 0.0629 \\ 
&  & MSE & 0.0054 & 0.0056 & 0.0051 & 0.0046 & 0.0041 \\ 
\cline{4-8}
& \multirow{3}{*}{0.10} 
& bias   & 0.0224 & 0.0198 & 0.0158 & 0.0126 & 0.0057 \\  
&  & sd  & 0.0641 & 0.0612 & 0.0619 & 0.0617 & 0.0618 \\  
&  & MSE & 0.0046 & 0.0041 & 0.0041 & 0.0040 & 0.0038 \\  
   \cline{1-8}
   \multirow{12}{*}{$\rho$}  
&   \multirow{3}{*}{-0.10}
& bias   & -0.0363 & -0.0250 & -0.0309 & -0.0136 & -0.0043 \\ 
&  & sd  & 0.0422 & 0.0541 & 0.0500 & 0.0265 & 0.0373 \\ 
&  & MSE & 0.0031 & 0.0035 & 0.0035 & 0.0009 & 0.0014 \\ 
\cline{4-8}
& \multirow{3}{*}{-0.05} 
& bias   & -0.0287 & -0.0065 & -0.0184 & -0.0121 & 0.0101 \\
&  & sd  & 0.0433 & 0.0589 & 0.0550 & 0.0283 & 0.0524 \\ 
&  & MSE & 0.0027 & 0.0035 & 0.0034 & 0.0009 & 0.0028 \\ 
\cline{4-8}
& \multirow{3}{*}{0} 
& bias   & -0.0306 & 0.0056 & 0.0096 & -0.0034 & 0.0100 \\ 
&  & sd  & 0.0417 & 0.0604 & 0.0575 & 0.0283 & 0.0403 \\ 
&  & MSE & 0.0027 & 0.0037 & 0.0034 & 0.0008 & 0.0017 \\ 
\cline{4-8}
& \multirow{3}{*}{0.05} 
& bias   & -0.0272 & -0.0136 & -0.0103 & -0.0110 & 0.0106 \\  
&  & sd  & 0.0433 & 0.0592 & 0.0585 & 0.0257 & 0.0503 \\ 
&  & MSE & 0.0026 & 0.0037 & 0.0035 & 0.0008 & 0.0026 \\ 
\cline{4-8}
& \multirow{3}{*}{0.10} 
& bias   & -0.0323 & -0.0228 & -0.0297 & -0.0110 & -0.0113 \\
&  & sd  & 0.0417 & 0.0544 & 0.0522 & 0.0275 & 0.0402 \\ 
&  & MSE & 0.0028 & 0.0035 & 0.0036 & 0.0009 & 0.0017 \\ 
   \cline{1-8}
  \multirow{12}{*}{$\eta$} 
   & \multirow{3}{*}{-0.1} 
   & bias & 0.0144 & 0.0168 & 0.0196 & 0.0202 & 0.0160 \\ 
&  & sd   & 0.0115 & 0.0102 & 0.0088 & 0.0077 & 0.0112 \\ 
&  & MSE  & 0.0003 & 0.0004 & 0.0005 & 0.0005 & 0.0004 \\ 
\cline{4-8}
& \multirow{3}{*}{-0.05} 
& bias   & -0.0032 & -0.0021 & -0.0041 & -0.0040 & -0.0010 \\ 
&  & sd  & 0.0149 & 0.0126 & 0.0105 & 0.0112 & 0.0110 \\ 
&  & MSE & 0.0002 & 0.0002 & 0.0001 & 0.0001 & 0.0001 \\ 
\cline{4-8}
& \multirow{3}{*}{0} 
& bias   & 0.0013 & 0.0012 & 0.0009 & 0.0006 & 0.0004 \\ 
&  & sd  & 0.0010 & 0.0012 & 0.0015 & 0.0015 & 0.0018 \\ 
&  & MSE & 0.0000 & 0.0000 & 0.0000 & 0.0000 & 0.0000 \\ 
\cline{4-8}
& \multirow{3}{*}{0.05} 
& bias   & -0.0013 & -0.0006 & 0.0030 & 0.0033 & 0.0034 \\ 
&  & sd  & 0.0118 & 0.0081 & 0.0082 & 0.0083 & 0.0089 \\ 
&  & MSE & 0.0001 & 0.0001 & 0.0001 & 0.0001 & 0.0001 \\ 
\cline{4-8}
& \multirow{3}{*}{0.10} 
& bias   & -0.0099 & -0.0112 & -0.0100 & -0.0090 & -0.0097 \\ 
&  & sd  & 0.0065 & 0.0054 & 0.0058 & 0.0072 & 0.0073 \\ 
&  & MSE & 0.0001 & 0.0002 & 0.0001 & 0.0001 & 0.0001 \\ 
\\[-1.8ex]\hline 
\hline \\[-1.8ex]
\end{tabular}}
\label{tab:rmse200}
\end{table}

\end{document}